# Magnetization switching by current in an elemental ferromagnetic single layer

Yanyan Yang & Weiwei Lin[*]

Key Laboratory of Quantum Materials and Devices of Ministry of Education, School of Physics, Southeast University, Nanjing 211189, China.

[*]e-mail: wlin@seu.edu.cn

**Current-induced magnetization switching, a fundamental phenomenon related to spin-transport of electrons, enables non-voltaic and fast information write, facilitating applications in low-power memory and logic devices. However, magnetization switching by spin-orbit torques is usually attributed to current flowing in the nonmagnetic metal layer of multilayers or in magnetic alloys with heavy elements. Here, we report perpendicular magnetization switching induced by current flowing in an elemental ferromagnet nickel single layer. This prototype structure demonstrates that current-induced magnetization switching is a general phenomenon of magnet. The results suggest that the current induces an effective transverse magnetic field with an out-of-plane component leading to the magnetization switching, different to the conventional spin-orbit torques. Our work opens the new insight and reveals the intrinsic mechanism of current-induced torques.**

In the 1970s, Berger et al. predicted in-plane current driven perpendicular magnetization reversal via magnetic domain wall motion[1-3]. Experimentally, in-plane current-induced deterministic switching of perpendicular magnetization has been reported in magnetic/nonmagnetic metal (NM) multilayers such as Pt/Co/Pt[4,5], and later, reversible current-induced magnetization switching (CIMS) has been observed in Pt/Co/AlO$_x$[6,7] and Ta/CoFeB/MgO[8] in which the switching polarity depends on the direction of the applied magnetic field along the current. The perpendicular magnetization switching is ascribed to spin-orbit torques (SOTs) from the spin Hall effect (SHE) in the adjacent NM layer with large spin-orbit coupling (SOC)[7,8] and/or the Rashba effect at the interfaces[6]. Various topological insulators (e.g. BiSbTe[9,10], Bi$_2$Se$_3$[11]), antiferromagnets (e.g. PtMn[12]) and van der Waals materials (e.g. WTe$_2$[13,14]) with large SOC have been utilized in SOT switching. However, the multilayers with a 4d transition metal Zr layer which are of weak SOC and small spin Hall angle have also been investigated similar CIMS, attributed to orbital Hall effect (OHE) in Zr[15,16].

CIMS has been also observed in single magnetic layers without the adjacent NM metal layer with SHE, such as non-centrosymmetric ferromagnetic semiconductor (Ga,Mn)As[17,18], ferromagnetic metals $L$10-FePt[19,20] and CoPt[21] alloys, rare earth ferrimagnetic CoTb[22,23] and FeTb[24] alloys, and non-collinear antiferromagnetic semimetal Mn$_3$Sn[25]. However, heavy elements with large SOC are included in most alloys. Low crystal symmetry[17-21], composition gradient[21,23,24] or locally asymmetric magnetic sublattices[13] have been suggested to be crucial for the switching behavior. CIMS in the thick single magnetic layer is mainly attributed to bulk spin torques due to large SOC[19,20] but the exact mechanisms remain unclear so far. Can perpendicular magnetization be switched by current in a single magnetic layer without large SOC ?

Here, we report the perpendicular magnetization switching induced by current flowing in a single elemental ferromagnetic nickel layer. This is the simplest and basic structure to realize CIMS without large SOC of heavy elements and absence of crystal asymmetry, composition gradient and magnetic sublattices. The result is crucial to verify the intrinsic mechanism of CIMS. We deposited the Ni films on undoped Si(100) substrates using magnetron sputtering at room temperature and patterned Hall bars (shown in Fig. S1) via photo-lithography and ion etching.

As shown in Fig. 1a, the current **I** is injected in the Ni layer along x axis and a magnetic field **H**$_x$ is applied parallel to the current. The perpendicular magnetization **M** of Ni can be switched by the injected current. Figure 1b shows the anomalous Hall resistance $R_{xy}$ loops with out-of-plane magnetic field $H_z$ of the device of 5.4 nm thick Ni layer and about 1 nm thick naturally oxidized NiO$_x$ on surface. The negative $R_{xy}$ refers to the Ni magnetization pointing to up, due to negative anomalous Hall angle of Ni. The $R_{xy}$ loops become square as the temperature $T$ decreases below 200 K, see also Fig. S2a. Both the remanence of the anomalous Hall resistance at zero out-of-plane field $R_{xy}^0/R_{xy}^s$ and the coercivity $H_c$ increase significantly with the $T$ decreasing below 200 K, as shown in Fig. S2b. This indicates the transition of the magnetic easy axis of the Ni layer from in-plane to out-of-plane around 200 K due to magnetoelastic anisotropy[26,27]. The magnetization remanence is 100% and the coercivity is 45 mT at 5 K. Figure 1c shows the current-induced switching in the Ni(5.4)/NiO$_x$(1) device at 50 K with the applied field $H_x$ along x axis. More current-induced switching loops for various $H_x$ at 50 K and 150 K are shown in Fig. S3. The anomalous Hall switching loop is counter-clockwise for the positive $H_x$, but is clockwise for the negative $H_x$. It indicates that the Ni magnetization favorites pointing to up (down) as the current is anti-parallel (parallel) to the field, as shown in Fig. 1a. The polarity of magnetization switching is the same as that of Pt/Co/AlO$_x$[6] but opposite to that of Ta/CoFeB/MgO[8]. The switching polarity may be related to current-induced domain wall motion[28,29] in Ni with a positive Dzyaloshinskii-Moriya intercation (DMI)[30]. No magnetization switching induced by current is observed as the applied field is along y axis.

**Temperature dependence of current-induced switching**

Figure 2a represents the current-induced switching loops of the Ni(5.4)/NiO$_x$(1) device under $\mu_0 H_x$ = 10 mT at several temperatures. The critical current $I_c$ and current density $J_c$ decreases by a factor of 10 with the $T$ increasing from 6 K to 200 K, as shown in Fig. 2b. $J_c$ is about 2 × 10$^7$ A cm$^{-2}$ at 200 K. This behavior is similar to that in single layers of magnetic alloys[18] and magnetic/NM multilayers[13,14], related to the reduce of the energy barriers and the propagation field of thermal excited magnetic domain wall motion as the $T$ increases[5]. Figure 2c shows the $H_x$ dependence of $I_c$ ($J_c$) at 6 K. The $J_c$ decreases as the magnitude of $H_x$ increases, similar to the behaviors in the magnetic/NM multilayers[6,9,12] and the magnetic alloys[20]. The in-plane magnetic field turns the in-plane orientation of the magnetic moments within the domain walls and induces asymmetric propagation of the domain walls[28-31]. One may see from Fig. 2c that the magnetization prefers pointing to up (down) as the current is anti-parallel (parallel) to the field. Figure 2d shows the $H_x$ dependence of current-induced switching ratio of magnetization $\Delta R_{xy}/2 R_{xy}^s$ at several temperatures. The switching ratio of magnetization increases and then decreases with increasing $H_x$, having a maximum at the field magnitude of about 10 mT at 6 K. This behavior is similar to the previous results in magnetic multilayers and alloys[12,20]. The increase of the switching ratio is due to the small in-plane field promotes the propagation of domain walls. Large in-plane field leads to the tilt of magnetization and reduces the perpendicular magnetization component for switching, resulting in the decrease of switching ratio. We have observed full magnetization switching induced by current in Ni at 6 K, as shown in Fig. S4. The maximum of current-induced switching ratio of magnetization increases with decreasing the $T$[32]. It is also resulted from the increase of the perpendicular magnetization remanence with decreasing the $T$ (see Fig. S2b) due to the enhancement of perpendicular magnetic anisotropy (PMA) at low temperatures.

**Thickness dependence of current-induced switching**

The $NiO_x$ on the surface is not antiferromagnetic, because no exchange bias is observed for the Ni/$NiO_x$ devices after in-plane field cooling down to 6 K, as shown in Fig. S5. In addition to Ni/$NiO_x$ we also measured the CIMS in devices of Ni layer with about 3 nm thick naturally oxidized $AlO_x$ layer. Figure 3a shows anomalous Hall resistance loops with out-of-plane fields in Ni($t_{Ni}$)/$AlO_x$(3) for several Ni thickness $t_{Ni}$ at 6 K. The Ni shows PMA for the Ni layer thickness from 5 to 8 nm. Figure 3b shows current-induced switching loop for several Ni layer thicknesses at 6 K under $\mu_0 H_x$ = 10 mT. It indicates the current switching ratio of magnetization is largest for the 6 nm thick Ni layer. Figure 3c shows the Ni layer thickness dependence of critical current density under $\mu_0 H_x$ = 10 mT at 6 K. The $J_c$ increases by 4 times with the Ni thickness from 5 nm to 8 nm, suggesting that the efficiency of current-induced torques is larger for the thinner Ni layer. This is similar to the interface dominated spin torques[33], but contrast to the bulk spin torques where $J_c$ almost does not change[22] or decreases[13] with the magnetic layer thickness increasing. This indicates that the interfaces of Si/Ni and Ni/$AlO_x$ play a crucial role in CIMS of Ni. As the Ni thickness decreases, the influence of the Si/Ni and Ni/AlO$_x$ interfaces dominates, leading to an enhancement in the efficiency of CIMS. The contributions of the Si/Ni and Ni/AlO$_x$ interfaces could be asymmetric. From the Ni thickness dependence of $J_c$, we suggest that CIMS does not mainly from the self-induced bulk spin torques in Ni layer. The CIMS in the Ni layer originates mainly from current-induced torques at the Ni/insulator interfaces.

**Current-induced effective magnetic field**

To further explore the mechanism of CIMS in the Ni, we quantitatively evaluate the current-induced effective magnetic field[5,29]. We measured the $R_{xy}$ vs $H$ loops in the Ni with various $I$ and $H_x$. Figure 4a shows the $R_{xy}$ loops with sweeping field $H$ at $\theta = 85°$ ($\theta$ is the angle of the applied field $H$ relative to the normal of the surface in yz plane, see Fig. S6a) in the Ni/NiO$_x$ at 14 K with $\mu_0 H_x$ = -8 mT. The $R_{xy}$ loops corresponding to positive (negative) currents shift to the left (right) under $\mu_0 H_x$ = -8 mT. There is no loop shift without the applied field along x axis, as shown in Fig. S6b. For the same current, the loop shift with $\mu_0 H_x$ = 8 mT is opposite to that with $\mu_0 H_x$ = -8 mT, as shown in Fig. S6c. As the current magnitude increases, the $R_{xy}$ loops become narrow. One may notice from Fig. 4a that with $\mu_0 H_x$ = -8 mT large positive (negative) current induces dip (peak) at low field of loop which does not appear for the small current. However, as shown in Fig. S6c, with $\mu_0 H_x$ = 8 mT large positive (negative) current induces peak (dip) at low field of loop. It suggests the roles of both the current and the applied field $H_x$ on the magnetization reversal process. The current dependence of the switching field is shown in Fig. 4b, where $H_L$ and $H_R$ refer to the switching field for field descending branch and ascending branch, respectively. The switching (coercive) field decreases remarkably with increasing the current, indicating current-induced torques and Joule heating helps the magnetization switching and thermal excited magnetic domain wall motion[4,5]. The $R_{xy}$ loop shift along the $H$ axis reverses with current, indicating the existence of current-induced effective magnetic field $H^{eff}$ which depends on the current direction. The current-induced effective magnetic field $H^{eff} = -H^{shift} = -½ (H_L + H_R)$, where $H^{shift}$ is the shift field of $R_{xy}$ vs $H$ loops. As shown in Fig. 4c, the magnitude of $H^{eff}$ increases with the current and saturated above 15 mA. The sign (direction) of $H^{eff}$ reverses with either the current direction or the applied field direction along x axis. This is very different to the Oersted field and the Rashba field. The $H^{eff}$ is close to +y (-y) direction as the current is anti-parallel (parallel) to the applied field along x axis. The current-induced effective field can be expressed as $\mu_0 H^{eff} = \chi J_c$, where $\chi$ is the coefficient of current-induced effective field. For the applied field of 8 mT along x axis, the coefficient of current-induced effective field $\chi$ in Ni is about $1.1 \times 10^{-7}$ mT A$^{-1}$cm$^2$ at 14 K.

To verify the direction of current-induced effective field, we measured the $R_{xy}$ vs $H$ loops in the Ni/NiO$_x$ at 14 K for several angles $\theta$ with various $I$. As shown in Figs. S6d-6f, the current-induced loop shift can also be observed at $\theta$ = 60°, 30°, and 0°. One may see from Fig. S6g that the sign of shift field $H^{shift}$ reverses with the current direction and the magnitude of $H^{shift}$ increases with $\theta$. The $H^{shift}(\theta)$ follows the $H^{shift}(0)/\cos\theta$ behavior, where $H^{shift}(0)$ is the shift field at $\theta$ = 0°, as shown in Fig. S6h. It indicates that the current-induced effective field is close to +y (-y) direction with a +z (-z) component as the current is anti-parallel (parallel) to the applied field along x axis. The current-induced effective field leads to the torques for the perpendicular magnetization switching.

To confirm the out-of-plane component of the current-induced effective field, we measured the $R_{xy}$ vs $H_z$ loops in the Ni/NiO$_x$ at 6 K with various $I$ and $H_x$. As shown in Fig. S7a, the current induces a clear shift of $R_{xy}$ loops in the Ni/NiO$_x$ with $\mu_0 H_x$ = 8 mT. It can be seen clearly that the $R_{xy}$ loops corresponding to positive (negative) currents shift to the right (left) under $\mu_0 H_x$ = 8 mT, and the shift direction reverses with $\mu_0 H_x$ = -8 mT, as shown in Fig. S7b. As the current increases, the loops become narrow and the perpendicular remanence magnetization decreases, indicating that the current reduces the energy barriers and the propagation field of thermal excited magnetic domain wall motion including the contributions of Joule heating[4,5,29]. The anomalous Hall loop shift along the $H_z$ axis reverses with current, indicating the out-of-plane component of current-induced effective field $H_z^{eff}$ which depends on the current direction. This is similar to the results in CoPt alloy[34]. The current dependence of the switching field $H_L$ and $H_R$ is shown in Figs. S7e and 7f for 8 mT and -8 mT, respectively. The switching (coercive) field decreases remarkably with increasing the current, indicating current-induced torques and Joule heating helps the magnetization switching and thermal excited magnetic domain wall motion[4,5]. The out-of-plane component of current-induced effective field $H_z^{eff} = - H_z^{shift} = -½(H_L + H_R)$, where $H_z^{shift}$ is the shift field of $R_{xy}$ vs $H_z$ loops. The magnitude of $H_z^{eff}$ is proportional to the current (in the measured current range) and the sign of $H_z^{eff}$ reverses with either the current direction or the applied field direction along x axis, as shown in Fig. S7g. The $H_z^{eff}$ points to up (down) as the current is anti-parallel (parallel) to the applied field along x axis. As shown in Figs. S7c and 7d, the current dependence of $R_{xy}^0$ at zero $H_z$ show a loop behavior and reverses for the opposite $H_x$, indicating that the current-induced effective field results in the perpendicular magnetization switching.

The applied magnetic field along x direction induces a tilt of out-of-plane magnetization and magnetization component along x direction within the chiral domain walls, as sketched in Fig. S8. Due to the inversion symmetry breaking by the substrate and surface, the effective field close to y direction with a z component can be induced by current around the chiral domain walls at the Ni/insulator interfaces with interfacial Dzyaloshinskii-Moriya interaction[28,29]. The current-induced effective magnetic field breaks the symmetry of up and down magnetization. The reversed magnetic domains expand due to the current-induced effective field, resulting in the perpendicular magnetization switching.

**Conclusion and outlook**

We have demonstrated the current induced perpendicular magnetization switching in an elemental ferromagnetic single layer, which is the basic structure to uncover the hidden mechanism of current-induced torques. The results prove that even for Ni single layer without large bulk SOC can achieve CIMS. This indicates CIMS does not require the NM layers with SHE and OHE, and the heavy elements in magnetic alloys with large SOC. Our work reveals that CIMS can be achieved without requiring SHE or OHE, nor the presence of heavy elements in magnetic alloys. This is not limited to specific materials and structures but represents a universal phenomenon in magnets. The results suggest that the current-induced effective transverse magnetic field at the Ni/insulator interfaces plays a dominant role in CIMS. This mechanism is intrinsic and general, applicable for the previous studied systems of magnetic multilayers and alloys. Our work suggests the CIMS can be realized without of complex fabrication processes for alloys and oxides, which is accessible for the integration. Enhancement of PMA at room temperature, the switching ratio free of field and reducing the critical current density for switching are essential to the applications of magnetic memory and logic devices based on current-induced torques.

**Methods**

**Sample fabrication and characterization**

The undoped Si(100) substrates were cleaned with hydrofluoric acid before the film deposition to remove the oxidizes on surface. The Ni films were deposited on the insulating Si substrates at room temperature using a dc magnetron sputtering with a base pressure of $7.8\times10^{-6}$ Pa and an Ar pressure of 0.3 Pa. By measuring the resistance of the film, we estimate that approximately 1 nm $NiO_x$ layer is on top by natural oxidation in air. For the Ni/$AlO_x$ devices, a 2 nm thick Al layer was sputtered on the Ni layer, followed by natural oxidation to form approximately 3 nm $AlO_x$ on top. The small angle X-ray reflectivity as shown in Fig. S1b indicates the small roughness of the Ni films. X-ray diffraction (XRD) shows the Ni film is (111) textured. The Ni films were patterned by UV photo-lithography and Ar ion milling into Hall bar devices with a stripe size of $20 \times 5$ $\mu m^2$, see Fig. S1a.

**Electrical transport measurements**

For the CIMS measurement, a pulsed electrical current with a duration of 1 ms was applied using a current source Keithley 6221. a d.c. current $I_{dc}$ of 0.1 mA was applied 1s after the current pulse and the Hall voltage $V_{dc}$ was recorded using a nanovoltmeter Keithley 2182A. The Hall resistance $R_{xy} = V_{dc}/I_{dc}$ was determined to characterize the magnetization state.

Current-induced anomalous Hall loop shift was measured to characterize the current-induced effective field. The anomalous Hall effect of the Ni devices was measured by sweeping the magnetic field and applying a constant field parallel to the current. A pulsed current with 1 ms duration was applied by Keithley 6221 and subsequently a dc current of 0.1 mA was applied 1s after the pulsed current to measure the Hall voltage.

**Acknowledgements**

We acknowledge support from the National Key Research and Development Program of China 2023YFA1406600.

**Author contributions**

W.L. conceived and supervised the work and developed physics model. Y.Y. fabricated the samples and performed the measurements. Both authors analyzed the data, discussed the results and wrote the manuscript.

**Competing interests** The authors declare no competing interests.

**Additional information**

**Correspondence and requests for materials** should be addressed to Weiwei Lin.

**Figures and legends**

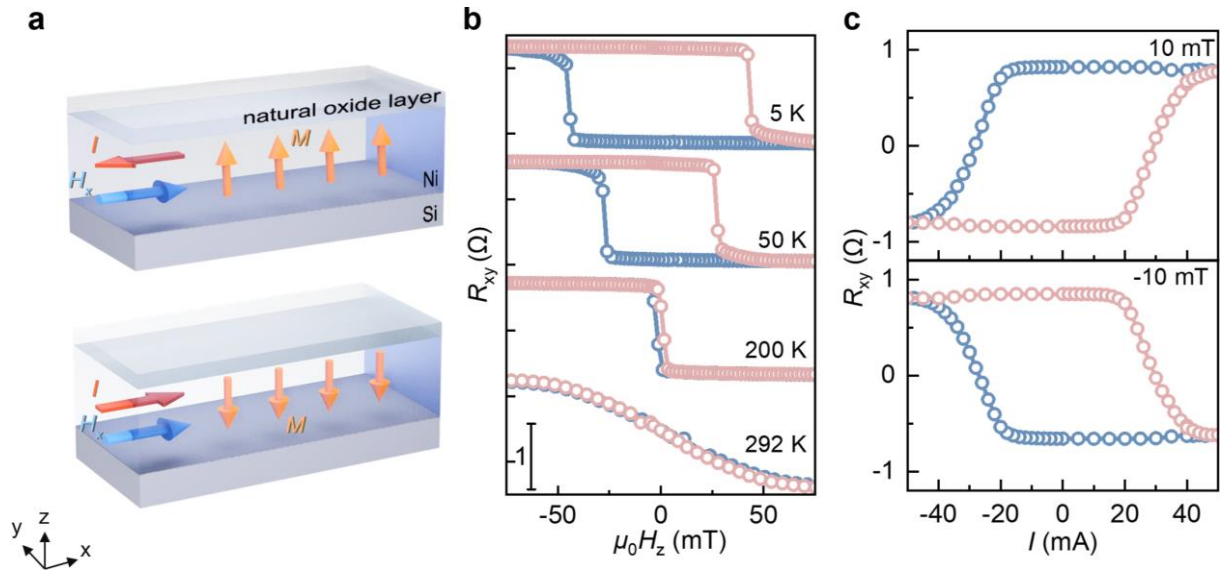

**Fig. 1 | Current-induced magnetization switching in Ni layer**. **a,** Schematic of current-induced magnetization switching in the Si/Ni/natural oxide layer structure. The magnetization is switched from up to down as the current is reversed from -x direction to +x direction, with a magnetic field applied along the +x direction. **b,** Anomalous Hall resistance loops with out-of-plane magnetic field at several temperatures for a Si/Ni(5.4)/NiO$_x$(1) device. **c,** Anomalous Hall resistance loops for the Si/Ni(5.4)/NiO$_x$(1) at 50 K by sweeping the pulsed currents under the applied magnetic field $\mu_0 H_x$ = 10 mT and -10 mT, respectively.

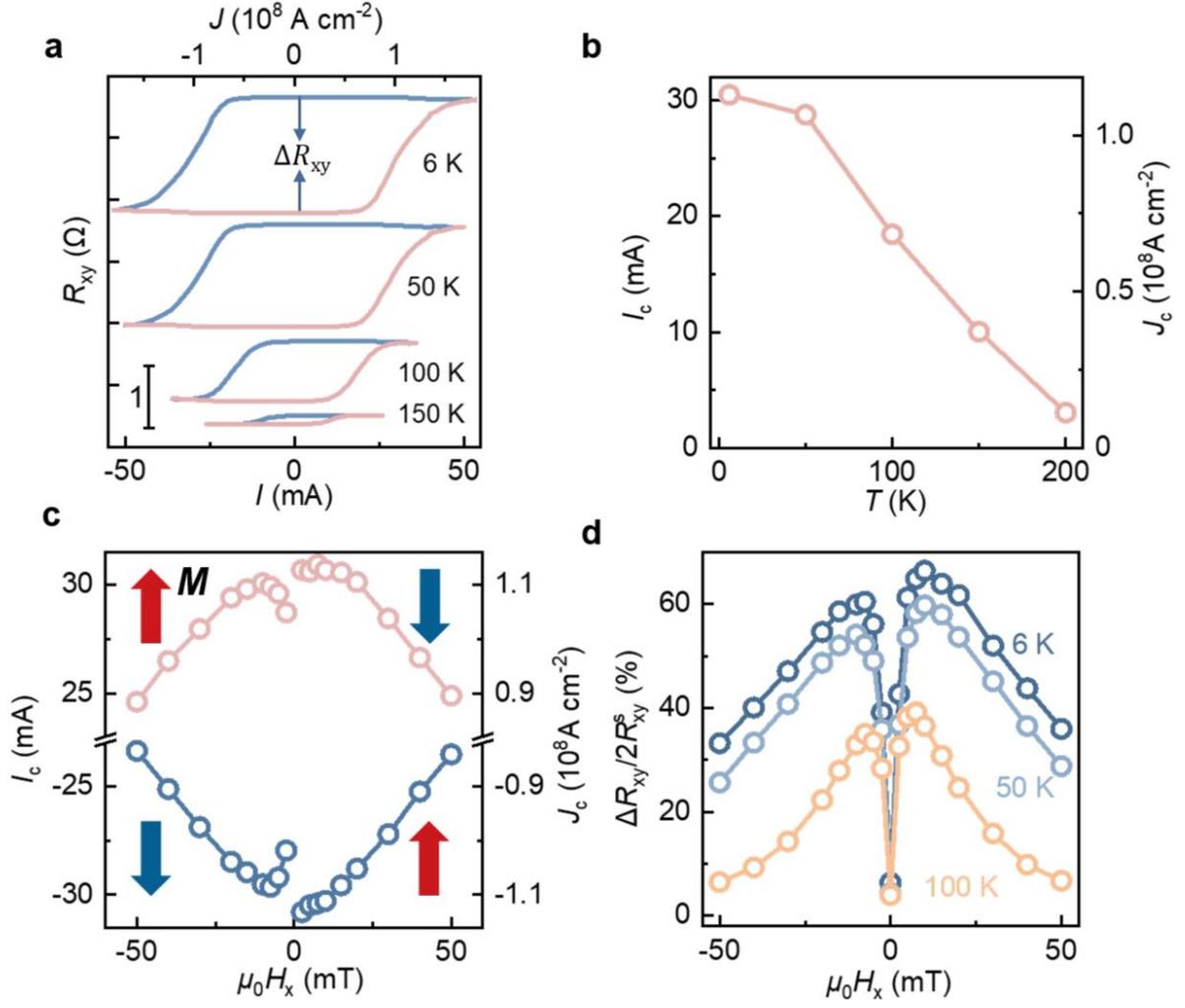

**Fig. 2 | Temperature and in-plane field dependences of CIMS in Ni layer. a,** Current-induced switching loops for Si/Ni(5.4)/NiO$_x$(1) at several temperatures under $\mu_0 H_x$ = 10 mT. **b,** Temperature dependence of critical current $I_c$ and critical current density $J_c$ at 10 mT. **c,** $H_x$ dependence of critical current $I_c$. **d,** $H_x$ dependence of current-induced switching ratio $\Delta R_{xy}/2R_{xy}^s$ at several temperatures.

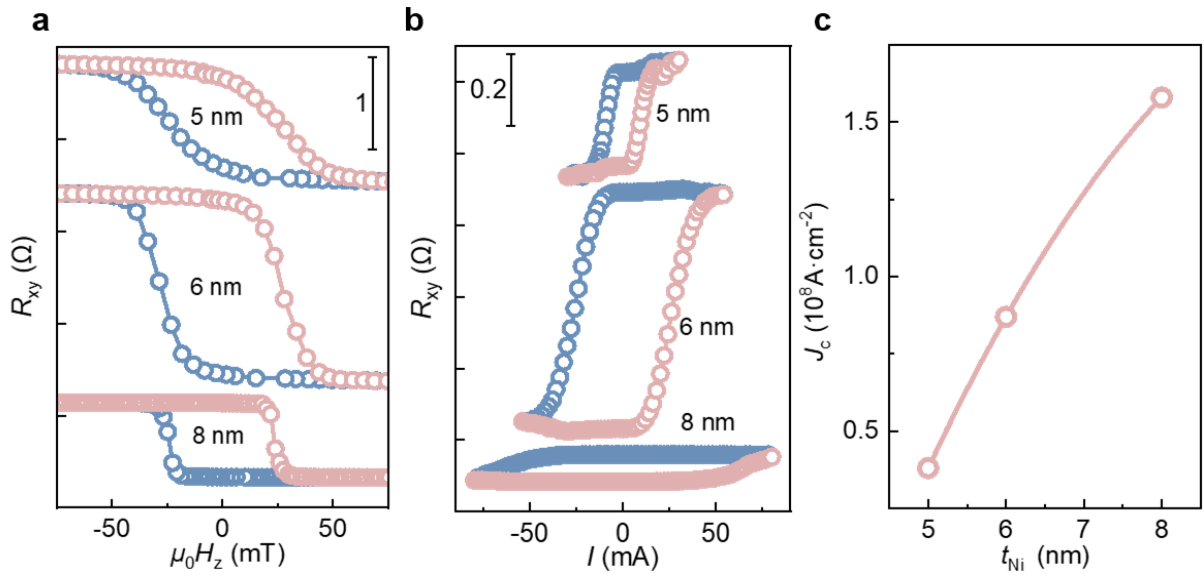

**Fig. 3 | Ni thickness dependence of CIMS. a,** Anomalous Hall resistance loops under out-of-plane magnetic field at 6 K for the Ni($t_{Ni}$)/AlO$_x$(3) devices with several Ni thickness $t_{Ni}$. **b,** Current-induced switching loops of Ni($t_{Ni}$)/AlO$_x$(3) with $\mu_0H_x$ = 10 mT at 6 K. **c,** Ni thickness dependence of critical current density with $\mu_0H_x$ = 10 mT at 6 K.

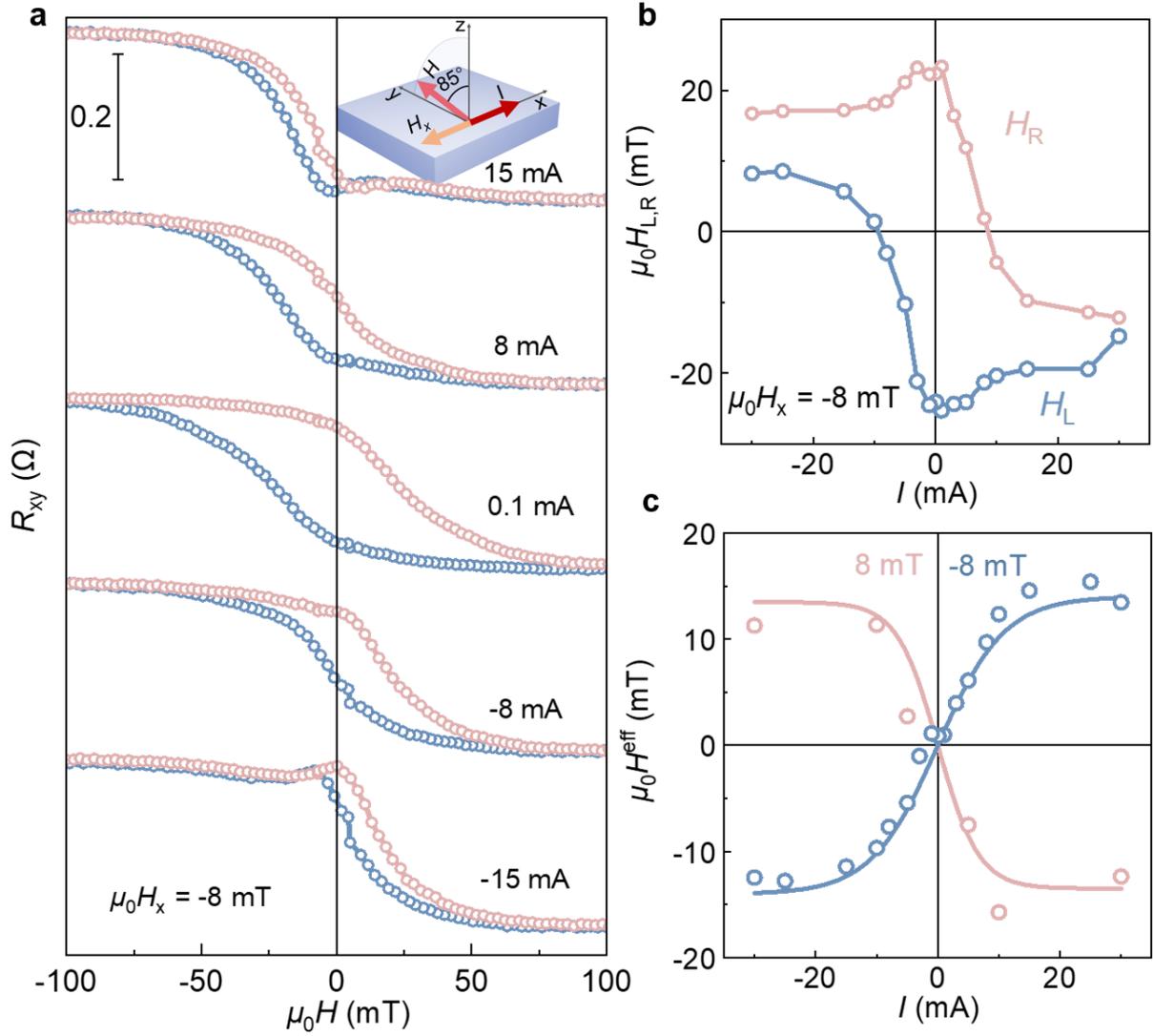

**Fig. 4 | Current-induced effective magnetic fields in Ni. a,** Anomalous Hall loops of Ni(5.4)/NiO$_x$(1) with varying fields at $\theta = 85°$ to the normal of surface in yz plane for several currents under $\mu_0H_x$ = -8 mT at 14 K. **b,** Current dependence of switching fields $H_L$ (field descending branch) and $H_R$ (field ascending branch) at $\theta = 85°$ with $\mu_0H_x$ = -8 mT. **c,** Current dependence of effective field $H^{eff}$ at $\theta = 85°$ with $\mu_0H_x$ = 8 mT and -8 mT, respectively.

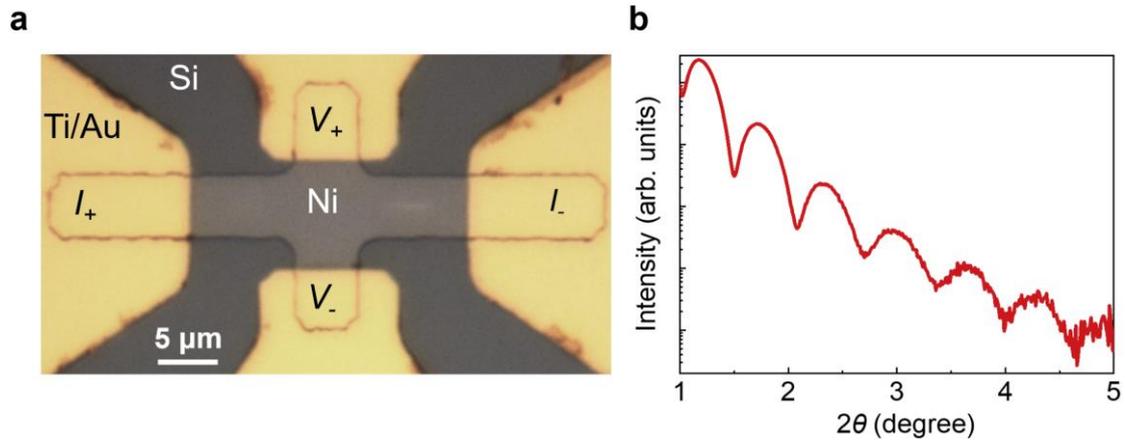

**Fig. S1 | Structural characterization of the Ni device. a,** Optical micrograph of the fabricated 20 μm × 5 μm Ni Hall bar. **b,** X-ray reflectivity of a Ni film.

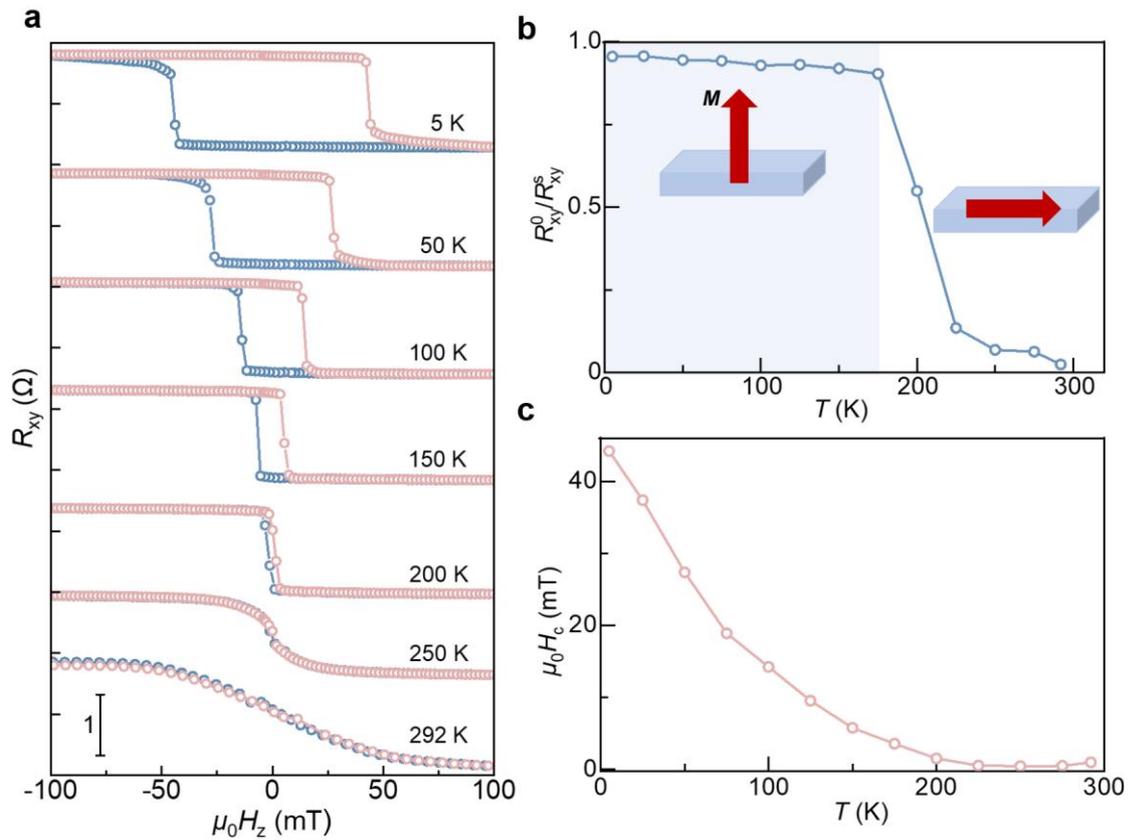

**Fig. S2 | Temperature dependence of anomalous Hall resistance and coercivity in Ni/NiO$_x$. a,** Anomalous Hall resistance loops of Ni(5.4)/NiO$_x$(1) with perpendicular fields at various temperatures. **b,** Temperature dependence of anomalous Hall remanence $R_{xy}^0 / R_{xy}^s$. **c,** Temperature dependence of coercive field.

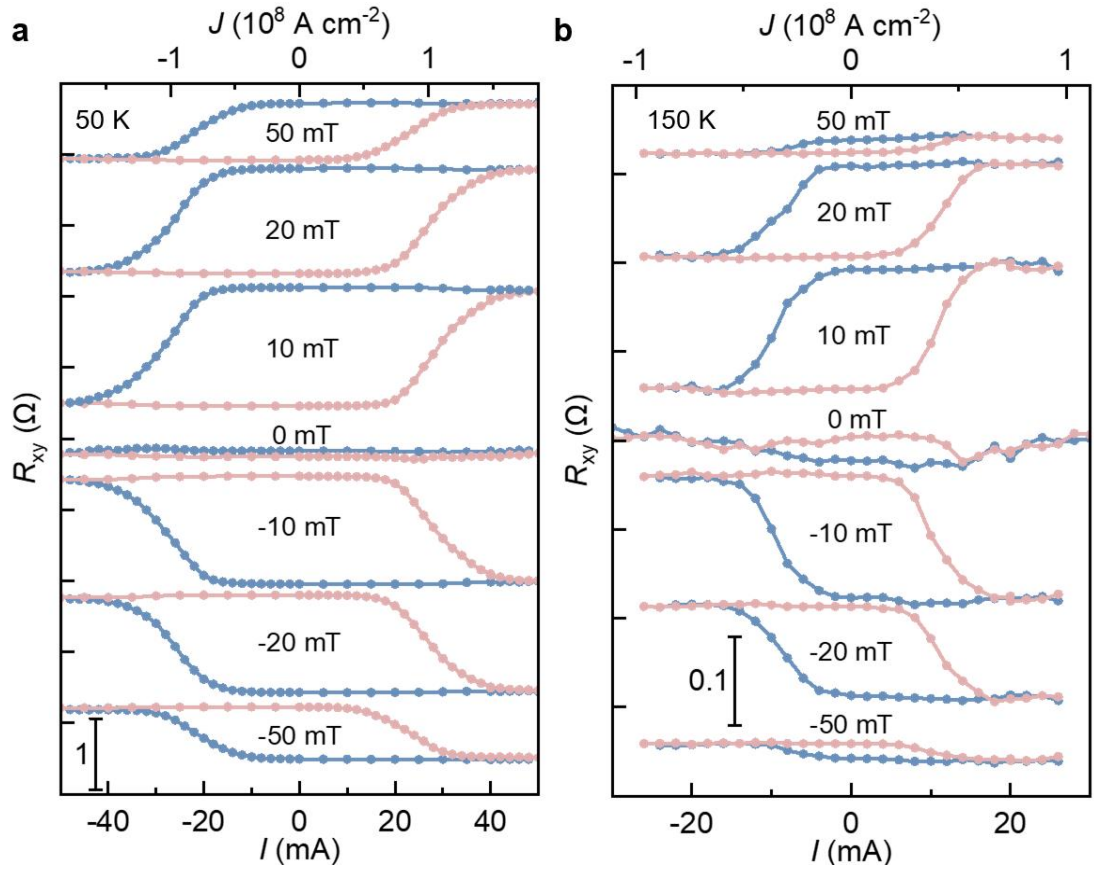

**Fig. S3 | CIMS loops of Ni for several in-plane fields.** Current-induced switching loops of Ni(5.4)/NiOx(1) for several $H_x$ at 50 K (**a**) and 150 K (**b**), respectively.

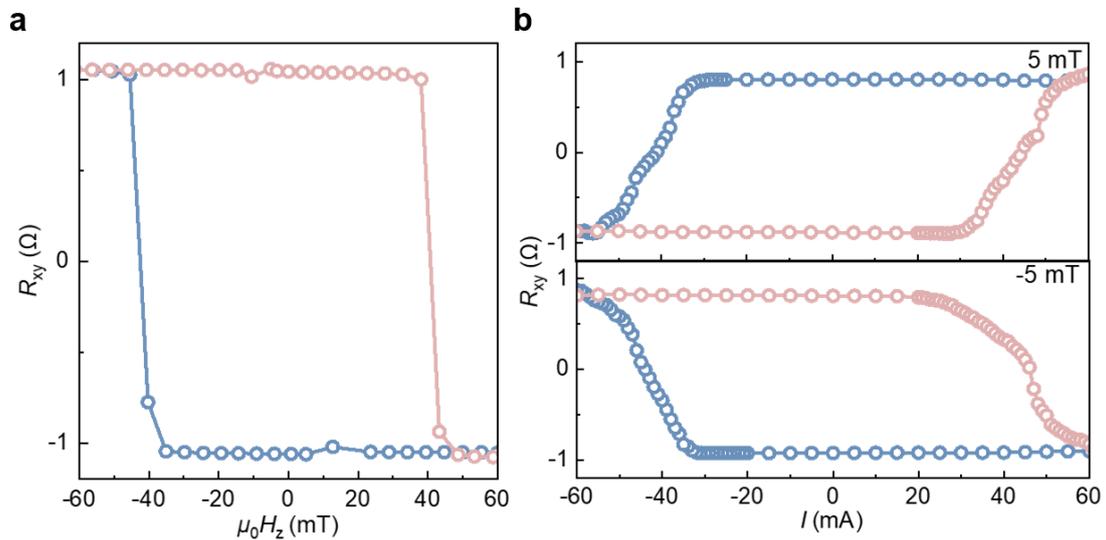

**Fig. S4 | Full magnetization switching induced by current in Ni. a,** Anomalous Hall resistance loop with perpendicular field at 6 K for Si/Ni(5.4)/NiO$_x$(1). **b,** CIMS loops for Si/Ni(5.4)/NiO$_x$(1) at 6 K with $\mu_0 H_x = \pm 5$ mT.

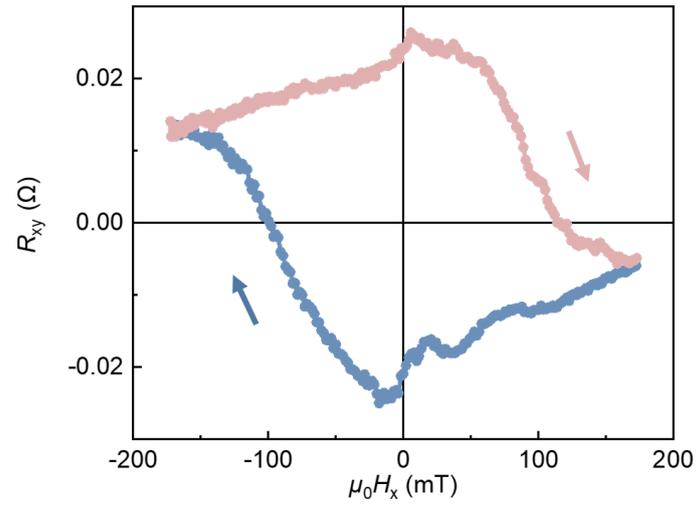

**Fig. S5 | Hall loop after field cooling.** Hall loop of Ni(6)/NiO$_x$(5) with magnetic field along x axis at 6 K and 0.1 mA after cooling under 200 mT field along +x direction. The absence of exchange bias indicates that the NiO$_x$ is not antiferromagnetic.

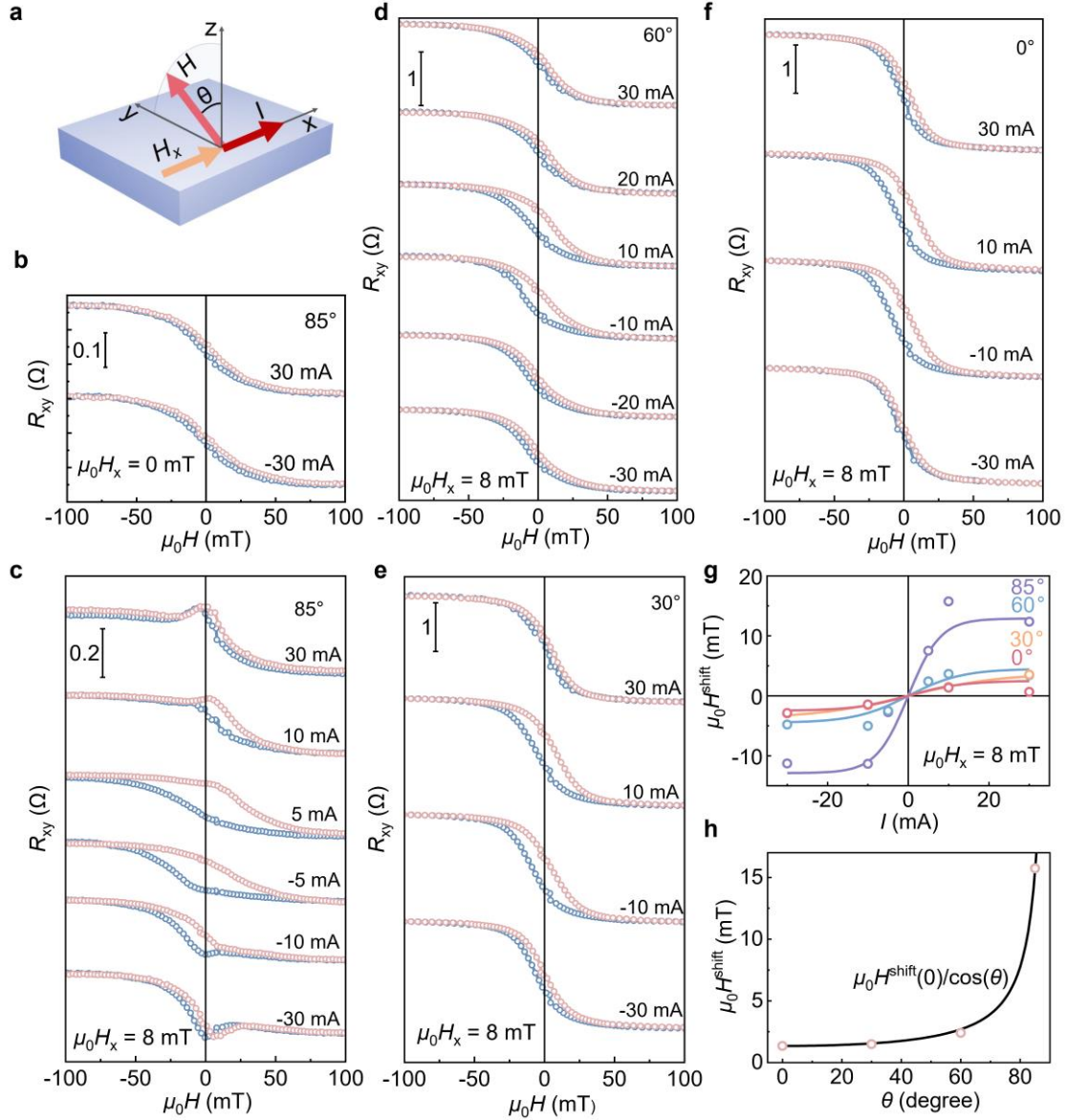

**Fig. S6 | Angle dependence of current-induced anomalous Hall loop shift at 14 K. a,** Schematic of the measurement of the current-induced anomalous Hall loop shift with sweeping the field at the angle $\theta$. $\theta$ is the angle of the applied field $H$ relative to the normal of the surface in yz plane. **b,** Anomalous Hall loops of the Ni(5.4)/NiO$_x$(1) with varying the field at $\theta = 85°$ for several currents without the $H_x$. **c, d, e, f,** Anomalous Hall loops of the Ni(5.4)/NiO$_x$(1) with varying the field at $\theta = 85°$ (**c**), 60°(**d**), 30°(**e**), and 0° (**f**) for several currents under an applied field $\mu_0H_x = 8$ mT. **g,** Current dependence of switching fields under $\mu_0H_x = 8$ mT at several angles $\theta$. **h,** Angle dependence of the $\mu_0H^{\text{shift}}$ under $\mu_0H_x = 8$ mT and $I = 10$ mA. The solid line represents the fitting to $\mu_0H^{\text{shift}}(0)/\cos(\theta)$.

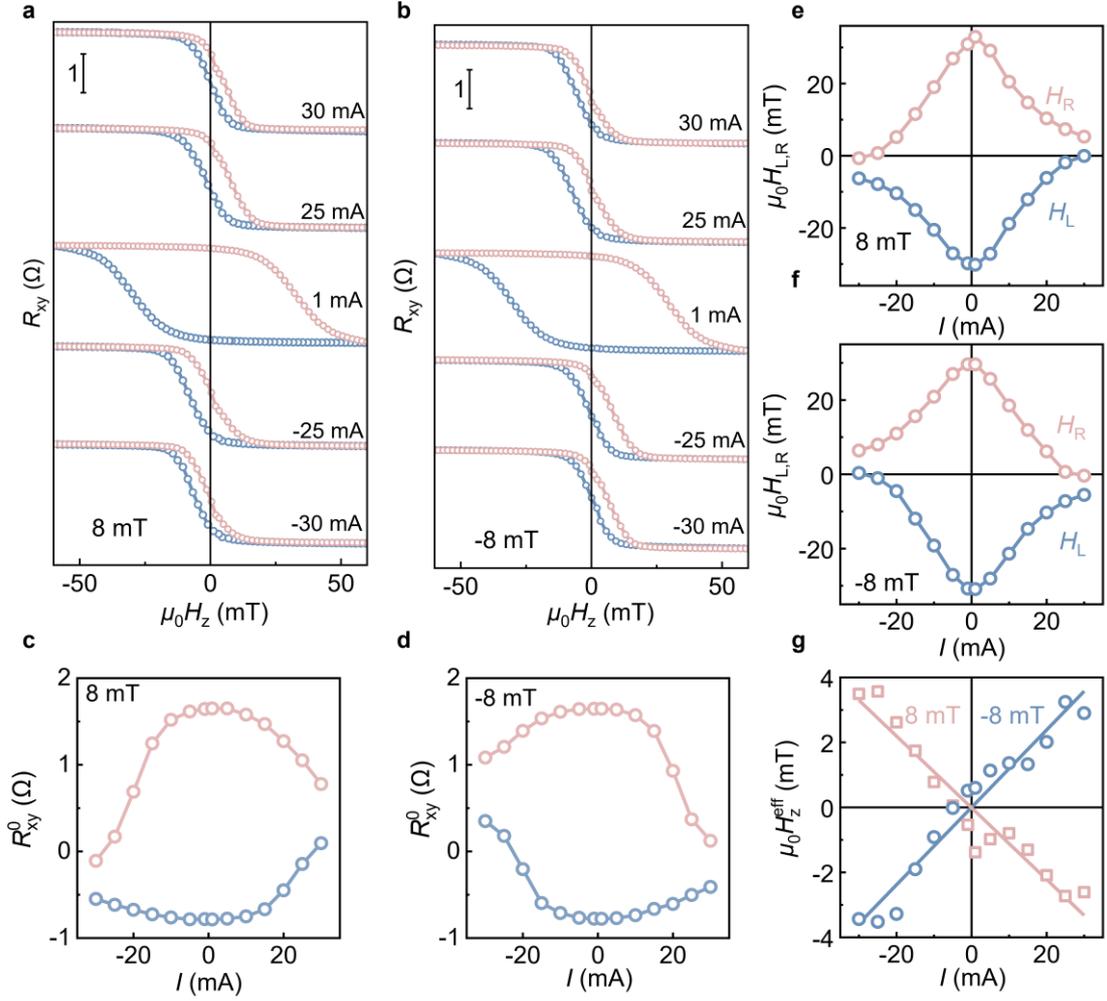

**Fig. S7 | Current-induced anomalous Hall loop shift with varying perpendicular field at 6 K. a, b,** Anomalous Hall loops of Ni(5.4)/NiO$_x$(1) with varying perpendicular field for several currents under an applied field $\mu_0H_x$ = 8 mT (**a**) and -8 mT (**b**), respectively. **c, d,** Current dependence of anomalous Hall resistance at zero out-of-plane field $R_{xy}^0$ with $\mu_0H_x$ = -8 mT (**c**) and 8 mT (**d**), respectively. **e, f,** Current dependence of switching fields $H_L$ and $H_R$ under $\mu_0H_x$ = 8 mT (**e**) and -8 mT (**f**), respectively. **g,** Current dependence of $H_z^{eff}$ with $\mu_0H_x$ = 8 mT and -8 mT, respectively. The solid line represents the linear fitting to the data.

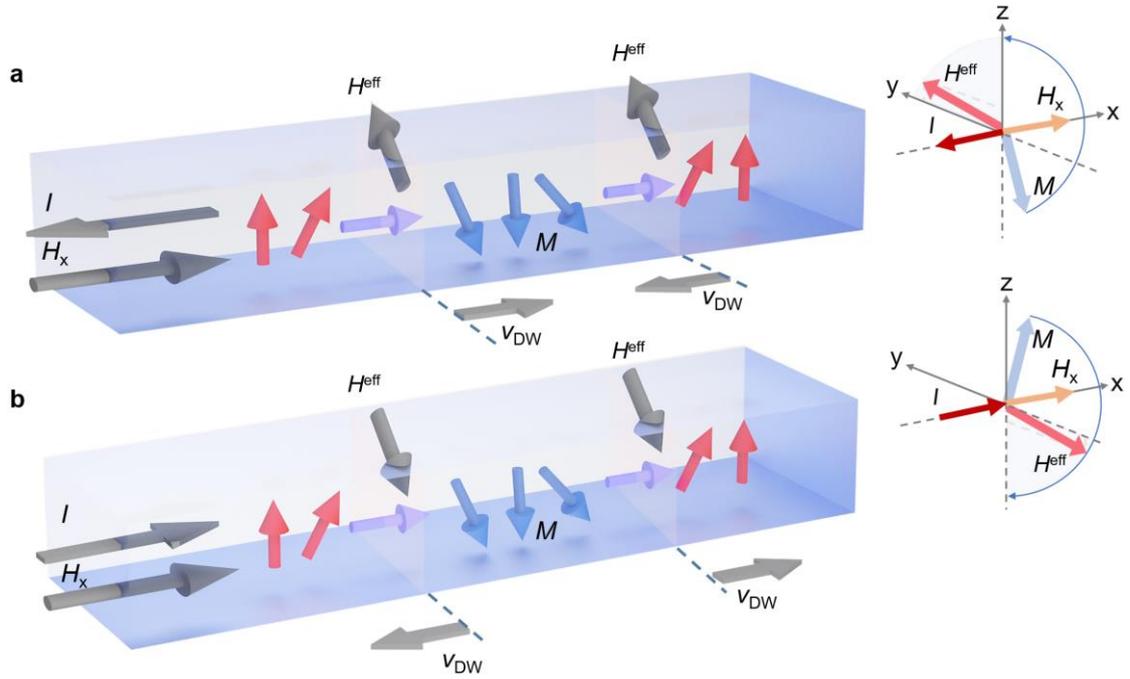

**Fig. S8 | Schematic of current-induced effective field. a,** The current-induced effective field $H^{\text{eff}}$ around domain walls points close to the +y direction with a +z component as the current is along -x direction with a magnetic field applied along the +x direction. The magnetization-up domains expand. **b,** $H^{\text{eff}}$ points close to the -y direction with a -z component as the current is along +x direction with a magnetic field applied along the +x direction. The magnetization-down domains expand.